\documentclass[iop]{emulateapj}
\usepackage{apjfonts}
\usepackage{graphicx}

\def\kms{${\rm km~s^{-1}}$ }
\def\sm{$M_{*}$}
\def\sigsm{$\Sigma_{*}$}
\def\sigSFR{$\Sigma_{\rm{SFR}}$}
\def\sigha{$\Sigma_{\rm{H}\alpha}$}
\def\rmxaa{RMxAA}
\def\ha{H$\alpha$}
\def\hb{H$\beta$}
\def\othree{[OIII] 5007}

\def\stwo{[SII] 6717+6731}
\begin{document}
\pagenumbering{gobble}

\title{SDSS-IV MaNGA: Spatially Resolved Star Formation Main Sequence and LI(N)ER Sequence}
\author{B. C. Hsieh\altaffilmark{1}, Lihwai Lin\altaffilmark{1},
J. H. Lin\altaffilmark{1,2}, 
H. A. Pan\altaffilmark{1},
C. H. Hsu\altaffilmark{1,2}, 
S. F. S\'{a}nchez\altaffilmark{3},
M. Cano-D\'{\i}az\altaffilmark{4}, 
K. Zhang\altaffilmark{5},
R. Yan\altaffilmark{5}, 
J. K. Barrera-Ballesteros\altaffilmark{6},
M. Boquien\altaffilmark{7}
R. Riffel\altaffilmark{8,9},
J. Brownstein\altaffilmark{10},
I. Cruz-Gonz\'{a}lez\altaffilmark{3}, 
A. Hagen\altaffilmark{11,12},
H. Ibarra\altaffilmark{3},
K. Pan\altaffilmark{13},
D. Bizyaev\altaffilmark{13,14},
D. Oravetz\altaffilmark{13},
A. Simmons\altaffilmark{13}}

\altaffiltext{1}{Institute of Astronomy \& Astrophysics, Academia Sinica,
P.O. Box 23-141, Taipei 106, Taiwan;
Email: bchsieh@asiaa.sinica.edu.tw}
\altaffiltext{2}{Department of Physics, National Taiwan University, 10617, 
Taipei, Taiwan}
\altaffiltext{3}{Instituto de Astronom\'{\i}a, 
Universidad Nacional Auton\'{o}ma de M\'exico, A.P. 70-264, 04510, 
M\'{e}xico, D.F., M\'{e}xico}
\altaffiltext{4}{CONACYT Research Fellow - 
Instituto de Astronom\'{\i}a, Universidad Nacional Aut\'onoma de M\'exico, 
Apartado Postal 70-264, Mexico D.F., 04510 Mexico}
\altaffiltext{5}{Department of Physics \& Astronomy,
177 Chem.-Phys. Building, University of Kentucky,
505 Rose Street Lexington KY 40506-0055, USA}
\altaffiltext{6}{Department of Physics \& Astronomy, 
Bloomberg Center for Physics and Astronomy,
Zanvyl Krieger School of Arts \& Sciences,
Johns Hopkins University, Baltimore, MD 21218, USA}
\altaffiltext{7}{Unidad de Astronom\'{i}a, 
Universidad de Antofagasta, Avenida Angamos 601, Antofagasta 1270300, Chile}
\altaffiltext{8}{Instituto de F\'{i}sica, 
Universidade Federal do Rio Grande do Sul,
Campus do Vale, Porto Alegre, RS, Brasil, 91501-970}
\altaffiltext{9}{Laborat\'{o}rio Interinstitucional de e-Astronomia, 
Rua General Jos\'{e} Cristino, 77 Vasco da Gama, 
Rio de Janeiro, Brasil, 20921-400}
\altaffiltext{10}{Department of Physics \& Astronomy,
University of Utah, Salt Lake City, UT 84112, USA}
\altaffiltext{11}{Department of Astronomy \& Astrophysics,
Pennsylvania State University, University Park, PA 16802, USA}
\altaffiltext{12}{Institute for Gravitation and the Cosmos,
Pennsylvania State University, University Park, PA 16802, USA}
\altaffiltext{13}{Apache Point Observatory and New Mexico State
University, P.O. Box 59, Sunspot, NM, 88349-0059, USA}
\altaffiltext{14}{Sternberg Astronomical Institute, Moscow State
University, Moscow}

\begin{abstract}
We present our study on the spatially resolved \ha~and \sm~relation
for 536 star-forming and 424 quiescent galaxies taken from the MaNGA survey.
We show that the star formation rate surface density (\sigSFR), 
derived based on the \ha~emissions, 
is strongly correlated with the \sm~surface density (\sigsm) 
on kpc scales for star-forming galaxies
and can be directly connected to the global star-forming sequence.
This suggests that the global main sequence may be 
a consequence of a more fundamental relation on small scales.
On the other hand, our result suggests that $\sim$ 20\% of quiescent galaxies
in our sample still have star formation activities in the outer region
with lower SSFR than typical star-forming galaxies.
Meanwhile, we also find a tight correlation between \sigha~and \sigsm~
for LI(N)ER regions, named the resolved `LI(N)ER' sequence, in quiescent galaxies, 
which is consistent with the scenario that LI(N)ER emissions
are primarily powered by the hot, evolved stars as suggested in the literature.

\end{abstract}


\section{Introduction}

It has been known for more than a decade that star-forming galaxies form
a tight sequence on the star formation rate and stellar mass plane,
the so-called `star formation main sequence' (SFMS)
\citep{bri04,noe07,dad07,pan09,kar11,whi12,spe14}.
The origin of the main sequence is often attributed to
the smooth mode of the star formation in galaxies
due to continuous accretion of the gas supply \citep{noe07}.
However, it has been challenging to reproduce the observed normalization
and slope of the main sequence in hydrodynamical simulations
and semi-analytical models \citep{dave08,dam09}.

One of the keys to understand the origin of the tight correlation
between SFR and \sm~is through the probe of these two quantities
on smaller scales,
i.e., the surface densities of SFR and \sm (\sigSFR~and \sigsm, respectively). 
If the relation between \sigSFR~and \sigsm~still holds on small scales,
it would suggest that the global SFR--\sm~relation is primarily
the outcome of the local correlation
and the mechanism that drives the star formation activity
with respect to the stellar mass could be universal
across various physical scales, similar to the situation where the well-known Kenicutt-Schmidt relation \citep{ken98a} between the star formation rate surface density and the cold gas surface density is found both locally \citep[e.g.,][]{big08} and globally. 
On the other hand, the lack of the correlation
between \sigSFR~and \sigsm~would otherwise suggest
a galaxy-wide process that regulates the SFR of galaxies as a whole.
Rosales-Ortega et al. (2012) and S{\'a}nchez et al.(2013) first show a tight correlation between
the stellar surface mass density and the surface star-formation density
for HII regions in nearby galaxies selected from
the Calar Alto Legacy Integral Field Area survey \citep[CALIFA;][]{san12}.
Later on, it has been found that at $z \sim 1$,
\sigSFR~in general traces the underlying \sigsm~\citep{nel16,wuy13}
using data from 3D-HST and CANDELS.
Recently, \citet{cano16} make a deeper analysis
using the CALIFA data and quantify the slope and amplitude of
the resolved \sigSFR~and \sm~relation on kpc scales for nearby galaxies
and showed that similar to $z \sim 1$ galaxies,
the spatially resolved relation also holds for star-forming galaxies
in the local Universe.

In this Letter,
we study the extinction-corrected \ha~surface density (\sigha) 
as a function of \sigsm~
for nearby galaxies taken from the MaNGA survey.
We not only double the sample size of local star-forming galaxies
compared to \citet{cano16},
but also extends the analysis to the quiescent population.
In addition, for the first time, we show that \sigha~
and \sigsm~form two separate sequences
for the HII and LI(N)ER regions in quiescent galaxies.
Throughout this paper we adopt the following cosmology:
\textit{H}$_0$ = 100$h$~\kms Mpc$^{-1}$ with $h$ = 0.7,
$\Omega_{\rm m} = 0.3$ and $\Omega_{\Lambda } = 0.7$.
We use a Salpeter IMF and the conversion from \citet{ken98b}:
SFR($M_\odot$ yr$^{-1}$) = $7.9 \times 10^{-42} L$(\ha) (ergs s$^{-1}$)
when deriving the star formation rate from \ha.

\section{Data and Sample Selection}
MaNGA is an IFU program to survey for 10,000 nearby galaxies 
with a spectral resolution varying from R $\sim$ 1400 
at 4000 \AA~to R $\sim$ 2600 at 9000 \AA~\citep{bla17,law15,yan16}. The targets are selected to represent the overall galaxy population with stellar masses greater than $10^{9} \rm M_{\odot}$ at $0.01 < z < 0.15$.
The angular size of each spaxel is 0".5 
while the average FWHM of the MaNGA data is 2".5, and
therefore, $\sim$ 20 neighboring spaxels are correlated.
The average number of spaxels is 800 per galaxy.
Our sample was drawn from the SDSS DR13 release \citep{alb17} containing 1392 galaxies, 
processed with the MPL-4 version of 
the MaNGA data reduction pipeline \citep{law15}. 
We adopt Pipe3D pipeline \citep{san16a} to perform the spectral line fitting, 
following the fitting procedures described in \citet{san16b}. 
The stellar continuum was first modeled with a linear combination 
of 156 single stellar population (SSP) templates 
with 39 ages and 4 stellar metallicities
that were extracted from the synthetic stellar spectra
from the GRANADA library \citep{mar05} and the MILES project \citep{san06}.
Then \sigsm~is obtained using the stellar populations 
derived for each spaxel.
To measure the emission line fluxes, 
we subtract the best-fit stellar continuum from the reduced data spectrum. 
The dust extinction of the emission line fluxes 
is corrected by using the Balmer decrement, 
adopting the Calzetti extinction law \citep{cal01} 
with $R_V = 4.5$ \citep{fis05}.

In this analysis, 
we confine our sample to galaxies that are not under interactions 
with other objects by excluding galaxies 
that have a spectroscopically-confirmed companion 
using the NSA catalog \footnote{http://www.nsatlas.org/data} 
or those identified as mergers identified 
by the Galaxy Zoo project \citep{dar10a,dar10b}.
The final sample consists of 1,085 galaxies.

Since the \ha~emission may be powered by various sources 
(e.g., star formation, AGN, old evolved stars, etc), 
we apply both the standard Baldwin-Phillips-Terlevich 
\citep{bal81,vei87,kau03,kew06}
excitation diagnostic diagrams and the WHAN diagram \citep{cid11}
to classify the emission line regions. 
More specifically, we adopt the \othree/\hb~versus \stwo/\ha~diagnostic 
with the dividing curves suggested in the literature 
\citep[e.g.,][]{kew01,kau03,cid10} to select `HII' and `LI(N)ER' regions,
and then use more conservative criteria
(i.e., log([NII]/{\ha})$ <-0.4$ and W$_{H\alpha}>5$\AA~
for HII regions and W$_{H\alpha}<3$\AA~for LI(N)ER regions,
where W is equivalent width) 
than those suggested in \citet{cid11}
to further clean our selections.

\begin{figure}
\epsscale{1.0}
\plotone{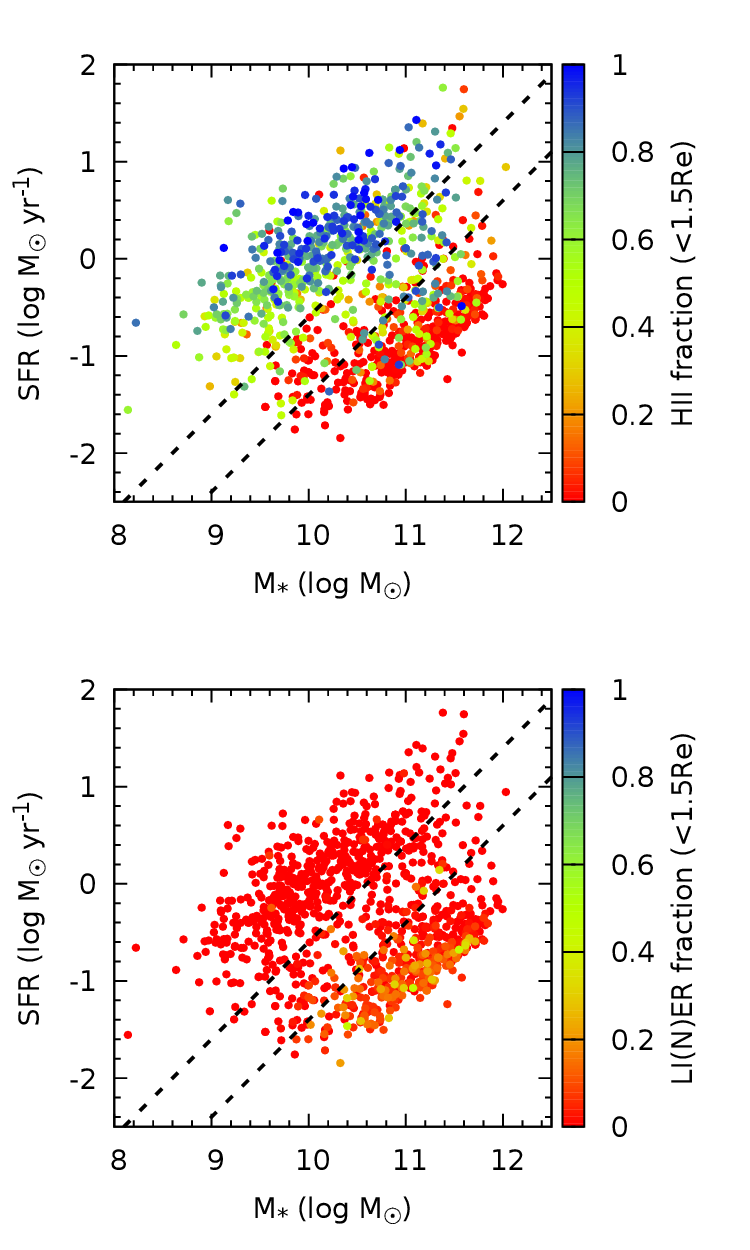}
\caption{Global SFR--\sm~relation with color-coded HII
and LI(N)ER fractions.
The black dashed-lines in both panels indicate two constand SSFRs;
-10.6 log(SFR M$_\odot^{-1}$) and -11.4 log(SFR M$_\odot^{-1}$).
See text for details.
\label{sfpplot}}
\end{figure}

To study the resolved star formation activity in each galaxy,
we define a quantity ``HII fraction'',
which is the ratio between the number of HII spaxels and
the number of spaxels within 1.5$\times$effective radius ($R_{\rm e}$)
which also meet at least one of the following criteria:
S/N(continnum) $>$ 3, S/N(\ha) $>$ 2, S/N(\hb) $>$ 2,
S/N([OIII]) $>$ 2, S/N([SII]) $>$ 2, and S/N([NII]) $>$ 2.
The HII fractions of our sample are color-coded and 
shown in the SFR--\sm~relation plot (the upper panel of Figure~\ref{sfpplot}).
The SFR and \sm~measurements are directly taken 
from the public MPA-JHU catalog 
\footnote{http://wwwmpa.mpa-garching.mpg.de/SDSS/DR7/}.
Since the Kroupa initial mass function (IMF) \citep{kro01} is adopted
in the MPA-JHU catalog while we use the Salpeter IMF \citep{sal55}, we add 0.2 dex to the MPA-JHU measured SFR and \sm~
to compensate for the differences.

There are two loci in this plot--the upper-left locus is the SFMS 
while the lower-right locus is referred to as the quiescent population.
As revealed in this figure, most galaxies in the SFMS have HII fractions
greater than $50\%$ (in green and blue),
while the quiescent sequence are dominated with 
galaxies with HII fractions less than $50\%$ (in orange and red). 
Similarly, we show the LI(N)ER fraction 
in the bottom panel of Figure~\ref{sfpplot}. 
It can be seen that quiescent galaxies have higher LI(N)ER fractions 
than star-forming galaxies, but none is higher than 0.6.

To select galaxies in both sequences for detailed analysis,
we apply two cuts with constant SSFRs
(shown as black dashed-lines in Figure~\ref{sfpplot}).
Galaxies with SSFR greater than -10.6 log(SFR M$_\odot^{-1}$) 
and those with SSFR less than -11.4 log(SFR M$_\odot^{-1}$)
are selected as the star-forming and quiescent populations, respectively. 
Galaxies between the two cuts are in the green valley
and will not be discussed in this letter.
After these selections,
there are 960 galaxies left for our analysis.

\section{Results}
\subsection{\sigha--\sigsm~relation for star-forming galaxies}

\begin{figure*}
\epsscale{1.0}
\plotone{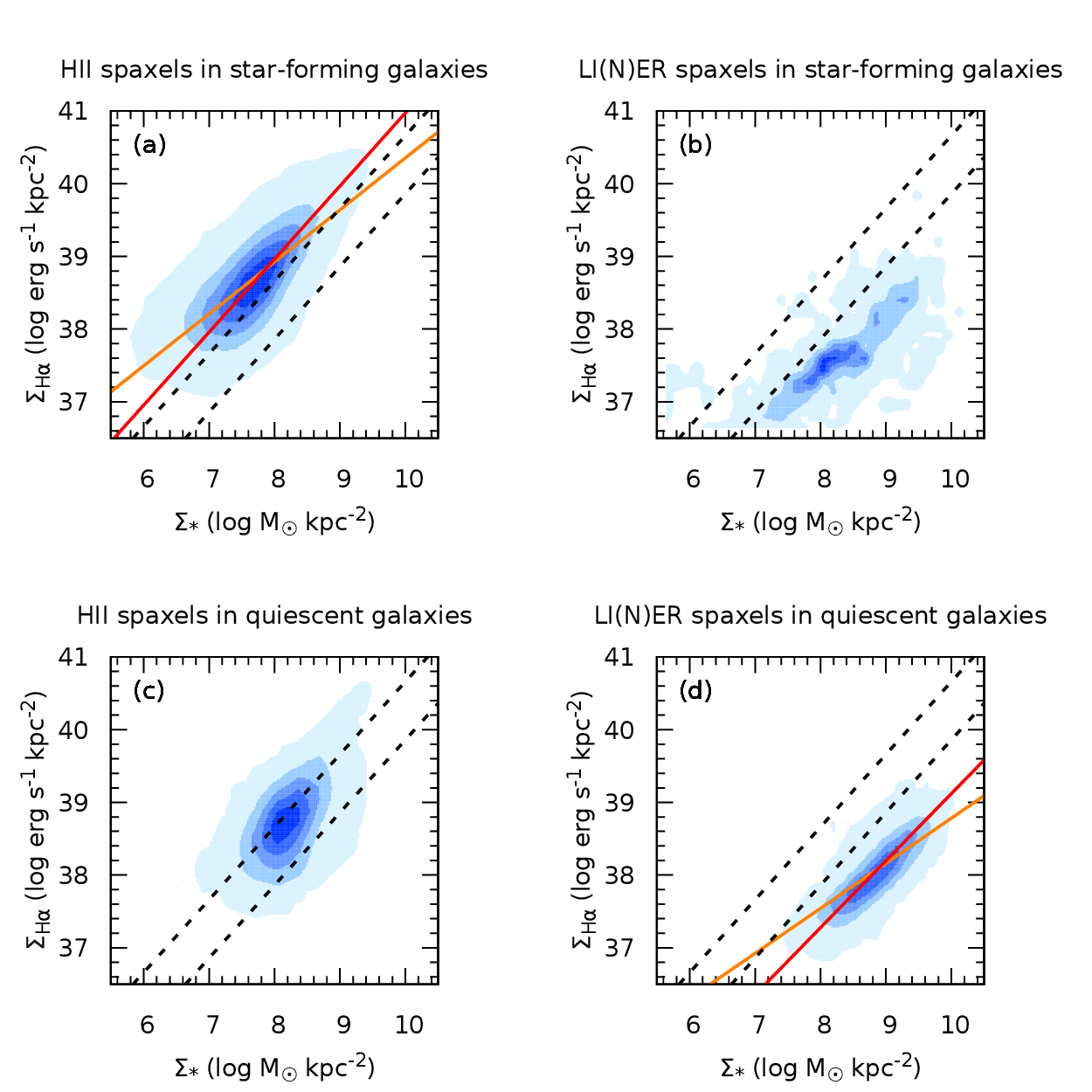}
\caption{The \sigha~--\sigsm~relations for HII and LI(N)ER spaxels
in star-forming and quiescent galaxies.
The blue color scheme of the contour 
indicates 1\%, 20\%, 40\%, 60\%, and 80\% of the peak density.
The two black dashed-lines are identical to those in Figure~\ref{sfpplot}.
The orange and red solid lines indicate
the best-fit lines using the OLS and ODR methods, respectively.
See text for details.
\label{allplot}}
\end{figure*}

To study the \sigha--\sigsm~relation for galaxies at different redshifts,
we need to compensate for the different physical size of spaxels.
Therefore, we convert the units of both \sigha~and \sigsm~ 
from per spaxel to per kpc$^{2}$, with inclination correction.
The inclination angle is derived using
cos$^2 i = ((b/a)^2 - \alpha^2) / (1 - \alpha^2)$,
where $i$ is the inclination angle, $\alpha$ is the intrinsic axis ratio,
and b/a is the observed ratio from the NSA catalog.
We use $\alpha = 0.13$ for our analyses.
The redshift information needed for this conversion 
is directly from the Pipe3D catalog.

The \sigha--\sigsm~relation of the HII spaxels of the star-forming galaxies
is shown in panel (a) of Figure~\ref{allplot}.
The two black dashed-lines indicate two constant SSFRs,
-10.6 log(SFR M$_\odot^{-1}$) and -11.4 log(SFR M$_\odot^{-1}$), respectively,
which are identical to those shown in the global SFR--\sm~relation plot
(Figure~\ref{sfpplot}), as references.
We perform a linear fitting with ordinary least square method (OLS).
The slope and zero-point (ZP) of the best-fit (orange solid line) are
$0.715\pm0.001$ and $33.204\pm0.008$, respectively.
We also perform the `orthogonal distance regression' (ODR) fitting,
which is to minimize the sum of the squares of
both the x residual and the y residual, thus is more suitable
for investigating the relation between two quantities.
By using the ODR fitting, the slope and zero-point of the best-fit 
(red solid line) are $1.005\pm0.004$ and $30.922\pm0.014$, respectively.
As shown in Figure 2, the ODR best-fit better represents 
the trend of the distribution than the OLS best-fit.
The fitting results using OLS and ODR are also summarized 
in Table~\ref{fittable}.
The values of zero-point in parenthesis are SFR densities
in the unit of log(M$_\odot$ yr$^{-1}$ kpc$^{-2}$).

We further compare our fitting results with \citet{cano16},
which study the spatially resolved SFR versus stellar mass relation 
for SFMS galaxies using data from CALIFA.
Although the definitions of star-forming and quiescent galaxies are
a bit different between \citet{cano16} and this letter,
our OLS fitting results are consistent with \citet{cano16}.
We also perform the ODR fitting for the data from \citet{cano16}
and the results are again consistent with each other.
The uncertainties of the fitting results in this letter 
are one order of magnitude smaller comparing to \citet{cano16}
because of the larger sample size of the MaNGA survey.

We repeat the same analysis for LI(N)ER spaxels
and the result is shown in panel (b) of Figure~\ref{allplot}.
At a given \sigsm, the \sigha for the LI(N)ER spaxels is lower than 
that for HII spaxels by more than one order of magnitude. 
As the number of LI(N)ER spaxels is too small 
for statistical meaningful analysis, 
we do not perform line fitting for this category.

\subsection{\sigha--\sigsm~relation for quiescent galaxies}

We next investigate the relation 
between the \ha~surface density and stellar mass surface density 
for the quiescent galaxies.
We repeat the same procedure for the HII/LI(N)ER spaxels as done for the SFMS,
except the inclination angle correction,
since it is non-trivial to estimate the inclination angles
of quiescent galaxies from the observed axis ratio.
The results are shown in panels (c) and (d) in Figure~\ref{allplot}.

Interestingly, a positive correlation between \sigha--\sigsm~ 
is also observed for HII spaxels but with lower SSFR by $\sim 0.5$ dex 
compared to the HII spaxels in star-forming galaxies. 
This offset is not totally unexpected 
since the quiescent population is either in the process of
undergoing quenching or has already experienced quenching,
which may alter the resolved \sigha--\sigsm~relation.

\begin{figure*}
\epsscale{1.0}
\plotone{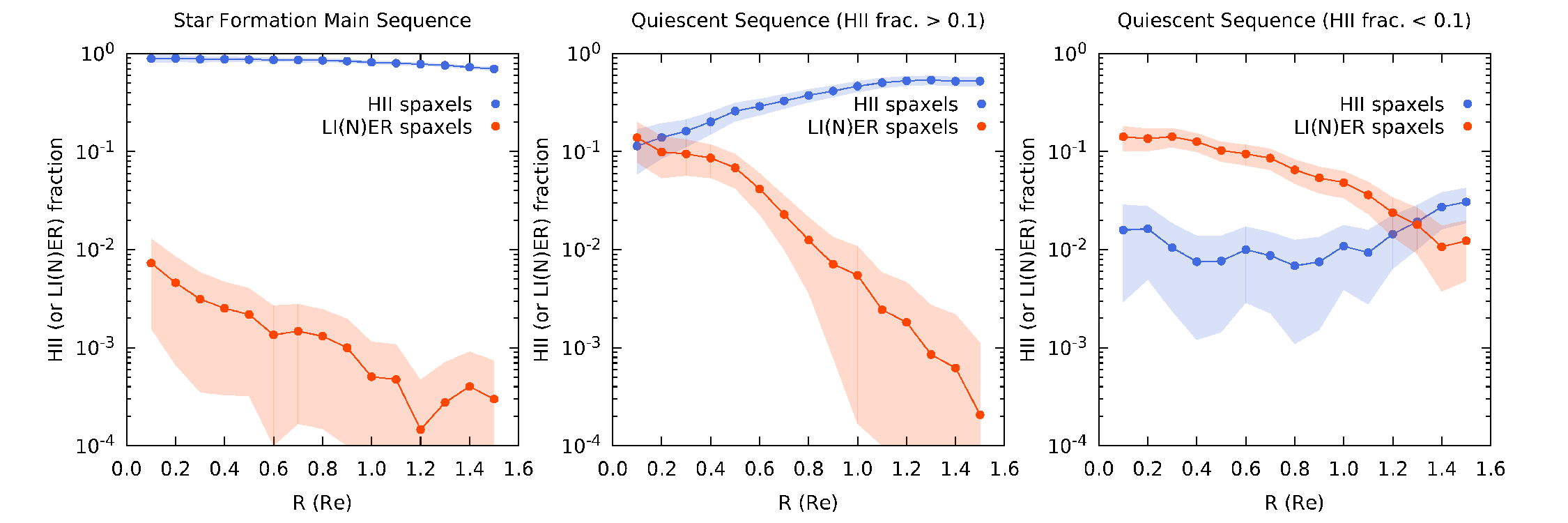}
\caption{Radial distributions of HII and LI(N)ER fractions.
Panels from left to right are for star-forming galaxies,
quiescent galaxies with HII fractions greater than 0.1, and
quiescent galaxies with HII fractions less than 0.1.
See text for details.
\label{galacto}}
\end{figure*}

To further investigate the properties of the HII spaxels in quiescent galaxies,
we make plots to show the radial distributions of the HII and LI(N)ER fractions
for both star-forming and quiescent galaxies.
The results are shown in Figure~\ref{galacto}.
We divide the sample into three catagories:
536 star-forming galaxies, 
92 quiescent galaxies with HII fractions greater than 0.1,
and 332 quiescent galaxies with HII fractions less than 0.1.
For a given spaxel, the radius is computed using
$R = \sqrt{x^2 + (y/cos\ i)^2}$ for a star-forming galaxy and
$R = \sqrt{x^2 + [y / (b/a)]^2}$ for a quiescent galaxy,
where $i$ is the inclination angle and $b/a$ is the axis ratio.
For star-forming galaxies, 
HII spaxels dominate from the core to outskirts, as expected.
For quiescent galaxies with high HII fractions, 
the HII fractions decrease toward the center
while the LI(N)ER fractions increase toward the center.
For quiescent galaxies with low HII fractions,
LI(N)ER fraction is higher than the HII fraction across the entire galaxies, 
except for the area beyond 1.4$R_{\rm e}$.
These galaxies are similar to cLIERs and eLIERs described in \citet{bel16}.

Most of the HII spaxels in quiescent galaxies shown in panel (c)
of Figure~\ref{allplot} belong to the quiescent galaxies 
with high HII fractions (e.g., the middle panel of Figure~\ref{galacto}).
Combining the results from the two figures suggests that 
$\sim$ 20\% of quiescent galaxies in our sample 
still have star formation activities 
in the outer region with lower resolved SSFR than typical star-forming galaxies.

Next we discuss the LI(N)ER spaxels.
The LI(N)ER fractions are much higher than the HII fractions
in quiescent galaxies, as shown in Figure~\ref{sfpplot}.
Although `LI(N)ERs' are often referred to 
low ionization nuclear emission line regions \citep{hec80}, 
previous works have shown that the 'LI(N)ER' region 
is not only confined in the galactic nuclei 
but can extend to several kpcs \citep{bel16,bel17}.
In addition to low activity AGNs, 
other possible ionizing sources of LI(N)ERs also include hot evolved stars 
\citep{bin94,yan12,cid11,sarzi10} and shocks \citep{dol95,dol15}. 
Studying the spatial distributions of LI(N)ER regions as well as 
how the strength of emissions depends on the stellar mass 
may shed lights on the origin of LI(N)ER emissions. 

Panel (d) in Figure~\ref{allplot} shows that
the LI(N)ER spaxels also form a very tight correlation 
between \sigha~and \sigsm.
We perform both the OLS and ODR fittings for the LI(N)ER distribution.
The slope and zero-point of the OLS best-fit (orange solid line) are
0.620$\pm$0.005 and 32.584$\pm$0.045, respectively.
The slope and zero-point of the ODR best-fit (red solid line) are
0.922$\pm$0.016 and 29.901$\pm$0.113, respectively.
The fitting results are summarized in Table~\ref{fittable}.

\begin{deluxetable}{lcc}
\tablecolumns{3}
\tablewidth{0pt}
\tablecaption{Fitting Results\label{fittable}}
\tablehead{
\colhead{} & \colhead{Star Forming} & \colhead{Quiescent} \\
\colhead{} & \colhead{(HII regions)} & \colhead{(LI(N)ER regions)} }
\startdata
Slope (OLS) & 0.715$\pm$0.001 & 0.620$\pm$0.005 \\
ZP (OLS) & 33.204 (-8.056)$\pm$0.008 & 32.584$\pm$0.045 \\
Scatter \tablenotemark{a} (OLS) & 0.159 & 0.095 \\
\\
Slope (ODR) & 1.005$\pm$0.004 & 0.922$\pm$0.016 \\
ZP (ODR) & 30.922 (-10.338)$\pm$0.014 & 29.901$\pm$0.113 \\
Scatter \tablenotemark{a} (ODR) & 0.127 & 0.071 \\
\enddata
\tablenotetext{a}{variance of residuals, defined as the weighted sum of squared residuals divided by degress of freedom}
\end{deluxetable}

Previously, \citet{sarzi10} has already shown that 
the \hb~flux closely traces the stellar continuum 
for early-type galaxies selected from the SAURON sample. 
The tight correlation we find between the \ha~surface density 
and stellar mass surface density for LI(N)ER spaxels, 
named `resolved LI(N)ER sequence', is thus not totally surprising. 
However, this is the first time we show that 
the \ha~emissions are directly correlated 
with the underlying stellar mass surface density. 
This can be consistent with the hot, evolved stars as the dominant mechanism 
powering the \ha~emissions in quiescent galaxies. 
A more detailed analysis comparing the properties of LI(N)ER regions 
between star-forming and quiescent galaxies will be presented 
in a forthcoming paper (Zhang et al. in prep.).

\subsection{Spatially resolved versus global \ha~(SFR)--\sm~distribution}

The correlation between the star formation rate surface density 
and stellar mass surface density on kpc scales seems to suggest that 
the star formation rate is controlled by the amount of old stars locally. 
Although the physical driver of this correlation remains unclear 
and is beyond the scope of this Letter, 
we note that there is a tentative evidence that 
the molecular gas surface density traces the stellar mass surface density 
on kpc scales (Lin et al. 2017), 
which may lead to the correlation 
between \sigSFR~and \sigsm~at a fixed star formation efficiency. 

To understand how the local relation is related to 
the global main sequence, we plot the resolved \sigSFR~-- \sigsm~ relation
with an axis-aligned sub-panel 
showing the global \ha~and \sm~relation in Figure 4. 
For the global distributions in the sub-panel, 
we convert the MPA-JHU derived SFR 
to the \ha~luminosity for both the star-forming and quiescent galaxies
with the same conversion.
As it can be seen, the global star-forming sequence is 
a continuous relation extended from the resolved \sigha~-- \sigsm~relation. 
The remarkable agreement between 
the resolved and global SFR and \sm~relations 
suggests that the global main sequence may originate 
from a more fundamental relation on small scales.  

On the other hand, the similarity between the resolved LI(N)ER sequence 
and the MPA-JHU global relation for the quiescent galaxies is rather suprising,
since the `SFR' in the latter catalog is calibrated 
using the relation between D4000 and sSFR based on emission-line galaxies. 

\begin{figure}
\epsscale{1.0}
\plotone{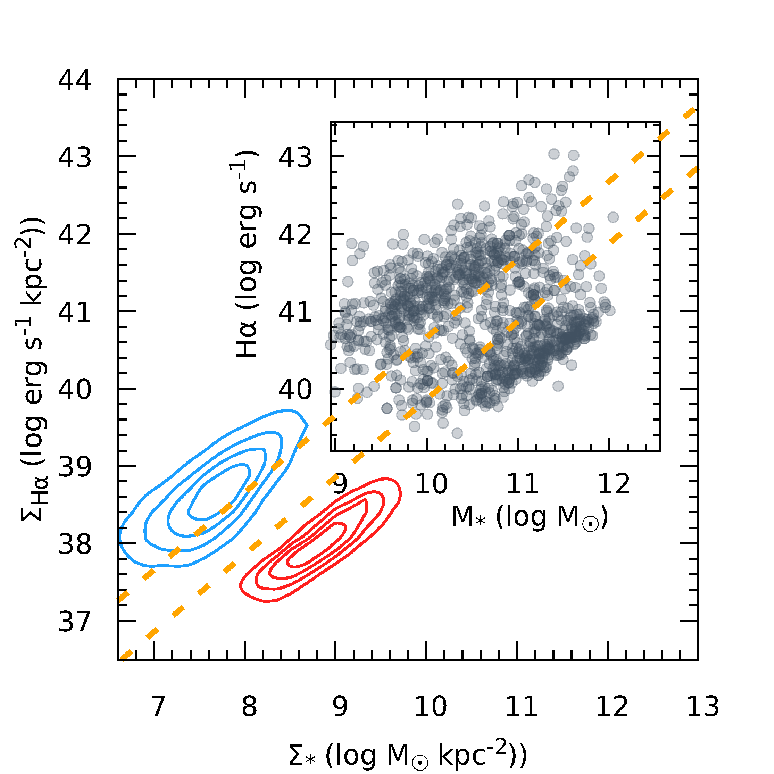}
\caption{Resolved and global \ha~and \sm~relation.
In the main panel, the spatially resolved distribution is shown in contours.
The blue and red contours indicate the distributions of 
the HII spaxels in the star-forming galaxies and 
the LI(N)ER spaxels in the quiescent galaxies, respectively.
The contour levels are 20\%, 40\%, 60\% and 80\% of the peak density
of the HII or LI(N)ER spaxels.
The two orange dashed-lines are identical 
to the black dashed-lines in Figure~\ref{sfpplot}.
The global distribution is shown in grey circles in the axis-aligned sub-panel.
See text for details.
\label{resolvedglobalplot}}
\end{figure}

\section{Conclusion}

In this letter, we study the global and resolved \ha~--\sm~relation
for the MaNGA MPL-4 sample.
We select star-forming and quiescent sequences 
with two constant SSFR criteria which consists of 960 isolated galaxies, 
and analyse the distributions of the HII and/or the LI(N)ER spaxels
for both populations in the \sigha~--\sigsm~plot.

We conclude our results below:

(1) There is a tight \sigha~--\sigsm~correlation of the HII spaxels 
for star-forming galaxies. The fitting results using the OLS method
are consistent with \citet{cano16}.

(2) The HII spaxels in quiescent galaxies have lower SSFR
than those in star-forming galaxies,
and the HII fractions decrease toward the center region
for quiescent galaxies with HII fractions greater than 0.1.
These results suggest that
$\sim$ 20\% of quiescent galaxies in our sample 
still have star formation activities 
in the outer region with lower SSFR than typical star-forming galaxies.

(3) The LI(N)ER spaxels in the quiescent galaxies show
a tight \sigha~--\sigsm~correlation.
This is the first time we show that 
the \ha~emission is directly correlated 
with the underlying stellar mass surface density
for regions classified as LI(N)ER.
This can be consistent with the hot, evolved stars as the dominant mechanism 
powering the \ha~emissions in quiescent galaxies.

(4) The global star-forming sequence is 
a continuous relation extended from the resolved \sigha~-- \sigsm~relation. 
The remarkable agreement suggests that the global main sequence may originate 
from a more fundamental relation on small scales.  

\acknowledgments
The work is supported by the Ministry of Science \& Technology 
of Taiwan under the grant MOST 103-2112-M-001-031-MY3 and 106-2112-M-001-034-.
This project also makes use of the MaNGA-Pipe3D data products.
We thank the IA-UNAM MaNGA team for creating it,
and the ConaCyt-180125 project for supporting them.
MB was supported by MINEDUC-UA project, code ANT 1655.
RR thanks to CNPq and FAPERGS for partial financial support.

Funding for the Sloan Digital Sky Survey IV has been provided 
by the Alfred P. Sloan Foundation, 
the U.S. Department of Energy Office of Science, 
and the Participating Institutions. 
SDSS-IV acknowledges support and resources from 
the Center for High-Performance Computing at the University of Utah. 
The SDSS web site is www.sdss.org. 
SDSS-IV is managed by the Astrophysical Research Consortium 
for the Participating Institutions of the SDSS Collaboration 
including the Brazilian Participation Group, 
the Carnegie Institution for Science, Carnegie Mellon University, 
the Chilean Participation Group, the French Participation Group, 
Harvard-Smithsonian Center for Astrophysics, 
Instituto de Astrof\'{i}sica de Canarias, The Johns Hopkins University, 
Kavli Institute for the Physics and Mathematics of the Universe (IPMU) 
/ University of Tokyo, Lawrence Berkeley National Laboratory, 
Leibniz Institut f\"{u}r Astrophysik Potsdam (AIP), 
Max-Planck-Institut f\"{u}r Astronomie (MPIA Heidelberg), 
Max-Planck-Institut f\"{u}r Astrophysik (MPA Garching), 
Max-Planck-Institut f\"{u}r Extraterrestrische Physik (MPE), 
National Astronomical Observatory of China, New Mexico State University, 
New York University, University of Notre Dame, 
Observat\'{a}rio Nacional / MCTI, The Ohio State University, 
Pennsylvania State University, Shanghai Astronomical Observatory, 
United Kingdom Participation Group, 
Universidad Nacional Aut\'{o}noma de M\'{e}xico, 
University of Arizona, University of Colorado Boulder, 
University of Oxford, University of Portsmouth, University of Utah, 
University of Virginia, University of Washington, University of Wisconsin, 
Vanderbilt University, and Yale University.

\end{document}